# Neural Network-Enhanced Frequency-Constrained Optimal Power Flow With Worst RoCoF and Frequency Deviation

Fan Jiang, *Student Member, IEEE*, Xingpeng Li, *Senior Member, IEEE*, and
Pascal Van Hentenryck, *Senior Member, IEEE*

***Abstract*—Frequency-constrained optimal power flow (FCOPF) should keep the rate of change of frequency (RoCoF) and the frequency deviation (FD) within their limits after a specified generator outage. The linearized frequency constraints commonly used in FCOPF are derived from aggregated system frequency response (SFR) models, which describe the center of inertia (COI) frequency with simplified governor dynamics and can underestimate the frequency excursions experienced by individual generators. To address this issue, this paper develops a neural network-based frequency predictor (NN-FP) trained on electromagnetic transient (EMT) simulations in PSCAD/EMTDC. The NN-FP predicts the worst RoCoF and the worst FD over the monitored generators, namely the generators at all buses other than the contingency bus. The trained NN-FP is reformulated and embedded into FCOPF with tightened frequency limits obtained from the test set prediction errors of NN-FP, leading to the proposed NN-FCOPF. The proposed model is compared with a traditional optimal power flow and a linearized FCOPF. EMT validation of 24-hour dispatches on the IEEE 9-bus and 39-bus systems shows that NN-FCOPF satisfies both limits in all hourly cases for the specified contingency, whereas the linearized FCOPF violates the limits in some hours because it underestimates the worst generator RoCoF and FD.**

***Index Terms*—Frequency deviation, neural network, optimal power flow, rate of change of frequency, real-time economic dispatch, system frequency response, system inertia.**

## I. Introduction

Frequency-constrained optimal power flow (FCOPF) incorporates post-contingency frequency requirements into generation dispatch, allowing operating cost and frequency security to be considered together [1]-[3]. Adjusting the pre-contingency active power outputs of synchronous generators changes the power deficit associated with a generator outage and the headroom available for primary frequency response. FCOPF can therefore improve the subsequent frequency response through preventive redispatch. Its effectiveness depends on whether the frequency constraints used in the optimization accurately represent the response to the resulting dispatch.

During the inertial and primary response stages, system frequency response (SFR) is commonly characterized by the rate of change of frequency (RoCoF) and the frequency deviation (FD) [4], [5]. RoCoF measures the speed of the frequency excursion, whereas FD records its depth, i.e., the distance between the frequency nadir and the nominal frequency. Tractable analytical approximations are widely used to incorporate these dynamic quantities into operational scheduling. For RoCoF, the initial swing equation approximation relates the disturbance power deficit to the available synchronous inertia [6]. This aggregate representation describes the initial center of inertia (COI) response [7]; the worst RoCoF experienced at an individual location can be larger [8].

FD estimation additionally requires the evolution of turbine-governor response. Anderson and Mirheydar [9] introduced a classic low-order SFR model in 1990, obtaining an analytical frequency response by simplifying the dominant turbine and governor dynamics. Low-order SFR models and related approximations have subsequently been used to formulate frequency constraints in OPF and scheduling [3], [10], [11], while governor ramping approximations offer another tractable route to primary response constraints [2]. More detailed formulations incorporate governor differential-algebraic equations into OPF [12], and Bernstein-polynomial approximations represent frequency dynamics in unit commitment (UC) [13]. Aggregate SFR formulations are widely used because they reduce the computational burden of evaluating frequency trajectories during optimization. Nevertheless, their accuracy depends on the aggregation and dynamic simplifications adopted. A constraint on a common system frequency does not explicitly bound the most severe excursion at every generator.

Machine learning offers an alternative by learning nonlinear frequency response mappings offline and embedding the trained predictors into scheduling models [14]. Sang et al. [15] developed conservative sparse neural approximations of nadir and stepwise frequency requirements for frequency-constrained UC, with measures to reduce optimistic prediction errors and extrapolation. She et al. [16] used neural networks to approximate FD and peak power in real-time virtual inertia scheduling. Both formulations include RoCoF and FD constraints, but retain an analytical RoCoF expression derived from an aggregate frequency response model. They therefore do not directly learn the worst local RoCoF from the simulated responses.

Spatial frequency variation has also been studied. The frequency divider formulation relates bus frequencies to generator speeds through the network [8] and provides a frequency estimation model. Tuo and Li [17] incorporated locational RoCoF constraints into security-constrained UC using network-coupled approximations and multiple measurement windows. Their learning-based formulation [18]

Fan Jiang and Xingpeng Li are with the Department of Electrical and Computer Engineering, University of Houston, Houston, TX 77204 USA (e-mail: fjiang6@uh.edu; xli83@central.uh.edu). Pascal Van Hentenryck is with the H. Milton Stewart School of Industrial and Systems Engineering, Georgia Institute of Technology, Atlanta, GA 30332 USA (e-mail: pascal.vanhentenryck@isye.gatech.edu).

instead predicts the highest locational RoCoF from simulation data. These two formulations constrain local RoCoF but do not jointly impose an explicit FD limit. Zhang et al. [19] learned the minimum bus frequency nadir and embedded it into preventive UC using region-of-interest active sampling, without jointly imposing a RoCoF constraint. Thus, [18] and [19] each learn a single locational indicator from simulation data, and none of these formulations jointly constrains the worst RoCoF and the worst FD, retains the unit dynamics and the spatial variation of an EMT model, and verifies the resulting dispatch in EMT.

Therefore, this paper proposes a neural network-based frequency predictor (NN-FP) that jointly predicts the worst RoCoF and the worst FD over a monitored set of generators from EMT simulation data, and embeds both frequency constraints in FCOPF. The predictor is embedded through an exact mixed-integer reformulation [20], [21], and empirical error margins are combined with electromagnetic transient (EMT) validation of the optimized dispatches. The main contributions are as follows:

- An NN-FP trained on EMT simulation data jointly predicts the worst RoCoF and the worst FD over the monitored set of generators. The targets retain the heterogeneous governor dynamics and the spatially nonuniform responses of the EMT model.
- The NN-FP is embedded in FCOPF through exact mixed-integer linear constraints, so that both frequency limits act on the dispatch simultaneously. Empirical margins based on held-out underprediction errors tighten both frequency limits and can mitigate the operational consequences of the prediction error.
- The resulting dispatches are validated in EMT simulations on the IEEE 9-bus and 39-bus systems against a traditional OPF and a linearized FCOPF with matched governor parameters, so that the final dispatches are checked on the model that generated the training data.

The remainder of this paper is organized as follows. Section II formulates the traditional OPF (T-OPF) and the linearized FCOPF (L-FCOPF) and discusses their limitations. Section III presents the proposed NN-FCOPF, including the offline generation of the EMT dataset and the training of the NN-FP, and the online reformulation of the trained predictor into the dispatch model. Section IV presents the case studies on the IEEE 9-bus and 39-bus systems, and Section V concludes the paper.

## II. Optimal Power Flow Formulations

This section formulates the two optimization models on which the proposed method is built and benchmarked, namely the traditional OPF (T-OPF) and the linearized FCOPF (L-FCOPF).

### A. Traditional Optimal Power Flow

The objective of the OPF framework is to minimize the total operating cost, as expressed in (1), where a quadratic function is employed to represent the generator cost.

$$min \sum_{i\in\mathcal{G}}(c_{2,i}P_i^2 + c_{1,i}P_i + c_{0,i}) \tag{1}$$

where $c_{2,i}$, $c_{1,i}$, $c_{0,i}$ are generator cost coefficients; $P_i$ is the active power output of generator $i$; $\mathcal{G}$ denotes the set of generators.

The constraints of the T-OPF consist of nodal power balance, power flow relations, generator output limits, and thermal limits of transmission lines, which are given in (2) to (5), respectively.

$$\sum_{k\in\mathcal{K}_b} P_{k,b}^{\text{fbus}} + P_b^{\text{load}} + P_b^{\text{loss}} = \sum_{k\in\mathcal{K}_b} P_{k,b}^{\text{tbus}} + \sum_{i\in\mathcal{G}_b} P_i\,,\ \forall b \in \mathcal{B} \tag{2}$$

$$P_k = \frac{\theta_p - \theta_q}{x_k},\ \forall k = (p,q) \in \mathcal{K} \tag{3}$$

$$P_i^{\min} \le P_i \le P_i^{\max},\ \forall i \in \mathcal{G} \tag{4}$$

$$-P_k^{\text{thm}} \le P_k \le P_k^{\text{thm}},\ \forall k \in \mathcal{K} \tag{5}$$

where $\mathcal{B}$ and $\mathcal{K}$ denote the sets of buses and transmission lines respectively; $P_{k,b}^{\text{fbus}}$ and $P_{k,b}^{\text{tbus}}$ represent the active power flowing from and to bus $b$ on line $k$, where $K_b$ and $G_b$ are the sets of lines and generators connected to bus $b$; $P_k$ denotes the active power flow on line $k$, whose terminal buses $p$ and $q$ have voltage angles $\theta_p$ and $\theta_q$, and $x_k$ is the reactance of line $k$; $P_i^{\min}$ and $P_i^{\max}$ are minimum and maximum active power limits of generator $i$; $P_b^{\text{load}}$ is the active power demand at bus $b$; $P_b^{\text{loss}}$ is the share of the network losses allocated to bus $b$, which is a constant in the DC power flow model; $P_k^{\text{thm}}$ is the thermal limit of transmission line $k$.

### B. Linearized Frequency-Constrained Optimal Power Flow

Most Independent System Operators (ISOs) solve the OPF problem in real time at regular intervals, typically every 5 minutes, which places high requirements on computational efficiency. In addition, the SFR is governed by the dynamics of the turbines, governors and rotating masses, making direct evaluation with detailed frequency response equations computationally expensive. As a result, SFR models in FCOPF are commonly simplified through linearization.

For the linearized formulation, the center of inertia (COI) frequency is defined as

$$f_{COI} = \frac{\sum_{i\in\mathcal{G}_h} f_i H_i S_i}{\sum_{i\in\mathcal{G}_h} H_i S_i} \tag{6}$$

where $\mathcal{G}_h$ denotes the subset of $\mathcal{G}$ that remains connected after the outage, $f_i$ represents the rotor speed frequency of generator $i$, and $H_i$ denotes the inertia constant of generator $i$ in seconds on its rated capacity $S_i$ in MVA. The product $H_iS_i$ is the kinetic energy of generator $i$ at nominal speed, and every inertia sum in this section is weighted by rated capacity in this way.

*1) RoCoF calculation*: The aggregated SFR model uses the initial COI RoCoF after a generator outage to approximate the worst RoCoF [6], [9]. Based on this premise, the RoCoF can be approximated using the generator swing equation, as given in (7).

$$\Delta P_m - \Delta P_e = \frac{2\sum_{i\in\mathcal{G}_h} H_i S_i}{f_0}\frac{df_{COI}}{dt} \tag{7}$$

where $\Delta P_{\text{m}}$ and $\Delta P_{\text{e}}$ denote the changes in mechanical input power and electrical output power in MW, respectively; $f_0$ denotes the nominal system frequency.

For a short interval immediately following the disturbance, the mechanical input power $\Delta P_{\text{m}}$ is assumed to remain constant, while the electrical output power $\Delta P_{\text{e}}$ changes instantaneously. The initial COI RoCoF is therefore approximated as:

$$\frac{df_{COI}}{dt} = -\frac{f_0 P_d}{2\sum_{i\in\mathcal{G}_h} H_i S_i} \tag{8}$$

where $P_{\text{d}}$ denotes the scheduled pre-disturbance output power of the tripped generator in MW, which is a decision variable of the dispatch model, and the sum in (8) is taken over $\mathcal{G}_h$ and excludes the tripped generator.

Accordingly, the RoCoF constraint can be written as:

$$\frac{f_0 P_d}{2\sum_{i\in\mathcal{G}_h} H_i S_i} \le R_{lmt} \tag{9}$$

where $R_{\mathrm{lmt}}$ denotes the RoCoF security limit, which is a positive value.

*2) FD calculation*: The typical full-order SFR model for a large-scale power system dominated by reheat turbine governors is illustrated in Fig. 1(a), in which each unit is represented by a droop gain, a lead-lag compensator, two first-order lags and a reheat block [4]. To facilitate simplified estimation of frequency behavior, Anderson and Mirheydar proposed a low-order SFR model, as shown in Fig. 1(b), in which nonlinearities and all time constants except the dominant ones in the generating unit equations are neglected [9].

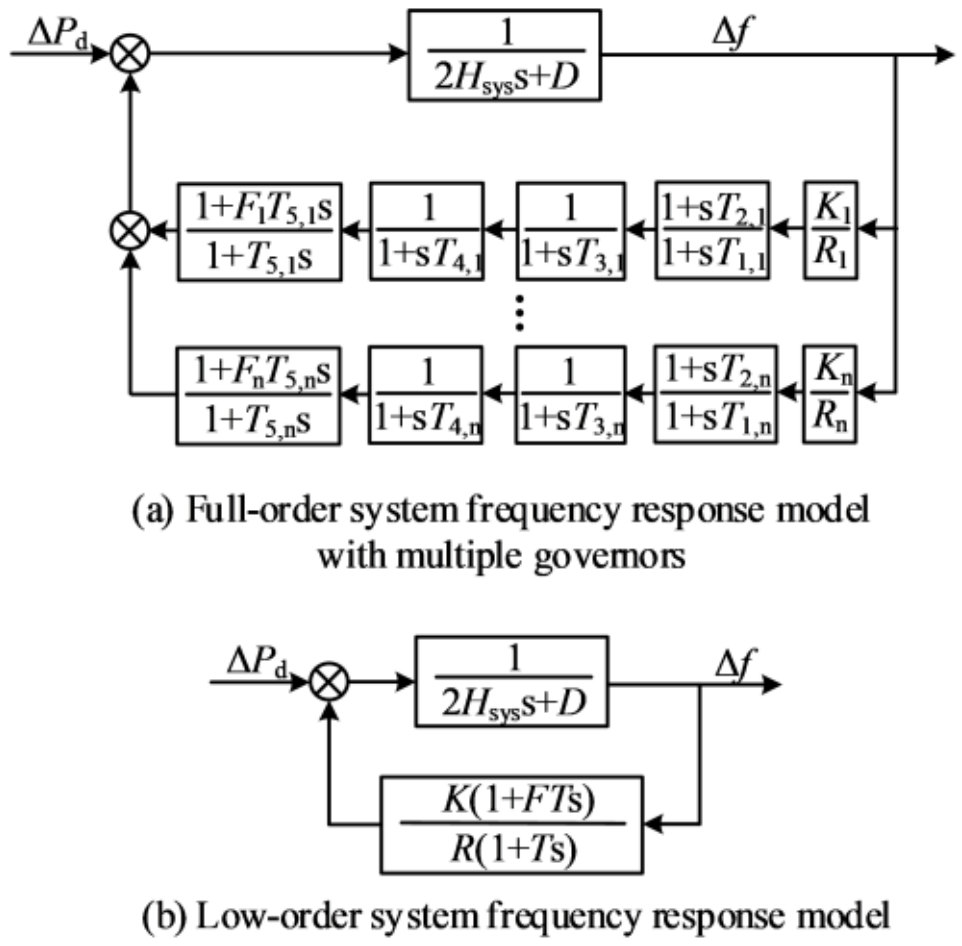

Fig. 1. System frequency response models of (a) full-order, and (b) low-order.

For a multimachine system in Fig. 1(a), $T_{1,\mathrm{i}}$ to $T_{5,\mathrm{i}}$ denote the time constants of generator $i$, $i = 1, 2, \ldots, \mathrm{n}$, $F_{\mathrm{i}}$ is the fraction of the total power generated by the high pressure turbine of generator $i$, $K_i$ is the mechanical power gain factor, and $R_i$ is the droop of generator $i$ in per unit on its rated capacity. $D$ denotes the system damping constant in per unit, and $H_{sys}$ stands for the system inertia constant in seconds; both are expressed on the total rated capacity of the generators that remain connected after the outage.

For the low-order system as in Fig. 1(b), the governor time constants are represented by a single aggregate time constant $T$ [4], [10]. Under a step power imbalance disturbance, i.e., $\Delta P_e(s) = \Delta P_d/s$, where $\Delta P_d$ is the lost generation $P_{\mathrm{d}}$ in per unit on the total rated capacity of the remaining generators, the resulting FD $\Delta f(t)$ in per unit of $f_0$ is given by (10).

$$\Delta f(t) = -\frac{\Delta P_d}{2H_{sys}T\omega_n^2} - \frac{\Delta P_d}{2H_{sys}\omega_d} e^{-\xi\omega_n t}[\sin(\omega_d t) - \frac{1}{\omega_n T}\sin(\omega_d t + \varphi)] \tag{10}$$

in which,

$$H_{sys} = \frac{\sum_{i\in\mathcal{G}_h} H_i S_i}{\sum_{i\in\mathcal{G}_h} S_i} \tag{11}$$

$$\omega_n = \sqrt{\frac{DR+K}{2H_{sys}RT}} \tag{12}$$

$$\omega_d = \omega_n\sqrt{1-\xi^2} \tag{13}$$

$$\xi = \frac{2H_{sys}R + T(DR+KF)}{2\sqrt{2H_{sys}RT(DR+K)}} \tag{14}$$

$$\varphi = sin^{-1}(\sqrt{1-\xi^2}) \tag{15}$$

$$F = \frac{\sum_{i\in\mathcal{G}_h} \frac{K_i F_i}{R_i} S_i}{\sum_{i\in\mathcal{G}_h} \frac{K_i}{R_i} S_i} \tag{16}$$

$$\frac{K}{R} = \frac{\sum_{i\in\mathcal{G}_h} \frac{K_i}{R_i} S_i}{\sum_{i\in\mathcal{G}_h} S_i} \tag{17}$$

$$T = \frac{\sum_{i\in\mathcal{G}_h} T_{5,i} S_i}{\sum_{i\in\mathcal{G}_h} S_i} \tag{18}$$

where $F$, $K$, $R$ and $T$ are the parameters of the low-order model aggregated from the full-order model, $\omega_{\mathrm{n}}$ stands for the natural frequency, $\omega_{\mathrm{d}}$ represents the damped natural frequency, $\xi$ denotes the damping ratio, and $\varphi$ is the phase angle associated with the damped oscillatory component of SFR.

Since only the dominant time constant $T_5$ among $T_1$-$T_5$ of the full-order model is retained in the low-order SFR model, the effective time constant $T$ is defined as the capacity-weighted average of $T_5$ across all participating generators. Likewise, (16) and (17) aggregate the governor gains of the remaining generators by rated capacity, so that $K/R$ is the equivalent regulation gain of the system and $F$ is the gain-weighted fraction of the turbine power. All sums in (11) to (18) are taken over $\mathcal{G}_h$ and exclude the tripped generator.

To determine the instant of the frequency nadir, the derivative of (10) is set to zero, as shown in (19).

$$\frac{d\Delta f(t)}{dt} = 0 \tag{19}$$

Assuming the disturbance occurs at time $t = 0$, the instant at which the frequency reaches its nadir can be expressed as (20).

$$t_{nadir} = \frac{1}{\omega_d} tan^{-1}\left(\frac{\omega_d T}{\xi\omega_n T - 1}\right) \tag{20}$$

The FD at $t_{nadir}$ is given by (21).

$$\Delta f = -\frac{R\Delta P_d}{DR+K}\left(1 + e^{-\xi\omega_n t_{nadir}}\sqrt{\frac{TK(1-F)}{2H_{sys}R}}\right) \tag{21}$$

Equation (21) is linear with respect to $\Delta P_{\mathrm{d}}$, and hence to $P_d$, and can therefore be incorporated into the FCOPF formulation through the FD constraint (22), which bounds the magnitude of the deviation in Hz.

$$f_0 \frac{R\Delta P_d}{DR+K}\left(1 + e^{-\xi\omega_n t_{nadir}}\sqrt{\frac{TK(1-F)}{2H_{sys}R}}\right) \le \Delta f_{lmt} \tag{22}$$

where $\Delta f_{\mathrm{lmt}}$ denotes the FD limit.

In summary, the L-FCOPF is the T-OPF augmented with the frequency constraints (9) and (22).

The L-FCOPF bounds the frequency response through (9) and (22), but both constraints describe the COI frequency of a single machine equivalent. Consequently, neither the RoCoF nor the FD experienced at an individual generator is constrained, and a dispatch that satisfies (9) and (22) may still trigger local frequency protection. Two sources of error are therefore present: the simplified governor representation, which affects the COI response itself, and the spatial variation of the frequency among the generators, which the COI does not capture. The case studies in Section IV show that the first source dominates the FD error in the 9-bus system, whereas the second dominates the RoCoF error in the 39-bus system. These limitations motivate the proposed NN-FCOPF, which replaces the analytical constraints with a data-driven predictor of the worst RoCoF and FD among the individual generators, as presented in Section III.

## III. Neural Network-Based Frequency-Constrained Optimal Power Flow

This section presents the proposed NN-FCOPF framework, which consists of an offline stage and an online stage, whose overall workflow is shown in Fig. 2.

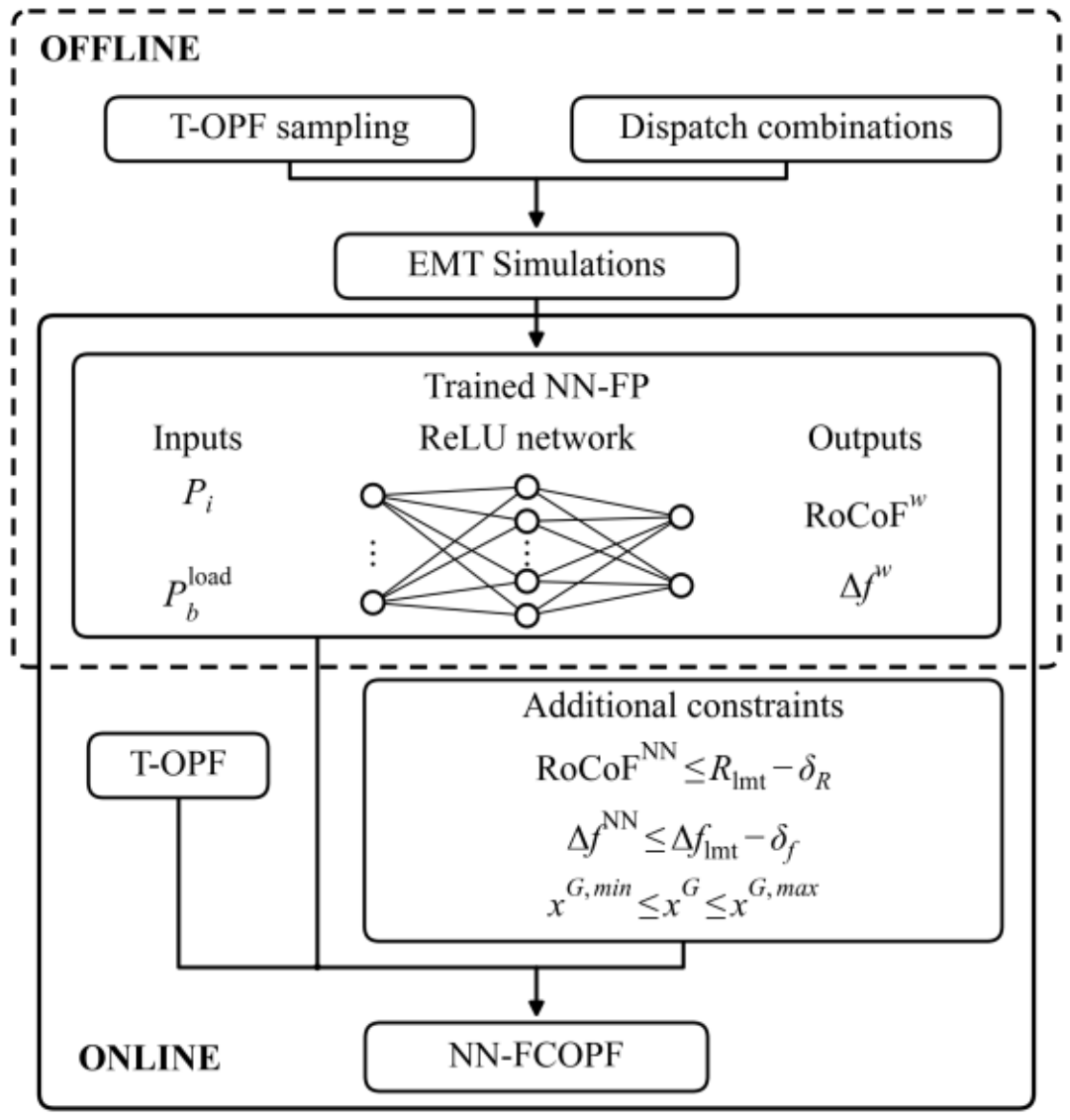

Fig. 2. Overall workflow of the proposed NN-FCOPF.

### A. Offline Stage: Dataset Generation and NN-FP Training

*1) Sampling of operating points*: Two sampling routes are combined to cover different operating conditions, as shown in the offline part of Fig. 2. The first route solves the T-OPF of Section II.A under varying load levels to obtain economic dispatches. These samples represent typical operating points, but their generator outputs are correlated through the merit order of the economic dispatch, which limits the diversity of generation allocations available for training. The second route supplements these points with dispatch combinations generated by Latin hypercube sampling. The load level and the output of the generator to be tripped are sampled within their admissible ranges, and the outputs of the remaining generators are allocated with randomly perturbed participation factors so that the active power balance and the output limits are satisfied. These samples decouple the size of the outage from the load level and broaden the coverage beyond economically optimal dispatches, which reduces the risk of the NN-FP learning spurious correlations that the optimizer could otherwise exploit.

*2) EMT simulation and labels*: Each sampled operating point is simulated in PSCAD/EMTDC under the specified generator outage at $t_0$. The generator active power setpoints $P_i$ and the bus active power demands $P_b^{\text{load}}$ form the input vector $x$, while the rotor speed frequency $f_i(t)$ of each generator recorded during the simulation provides the output labels. The two outputs of the dataset are defined as the worst RoCoF magnitude and the worst FD over the set $\mathcal{G}_m$ of monitored generators, which contains the generators at all buses other than the bus where the contingency occurs:

$$\text{RoCoF}^{\text{w}} = \max_{i\in\mathcal{G}_m} \max_{t\ge t_0} \frac{|f_i(t+T_{\text{w}}) - f_i(t)|}{T_{\text{w}}} \tag{23}$$

$$\Delta f^{\text{w}} = f_0 - \min_{i\in\mathcal{G}_m} \min_{t\ge t_0} f_i(t) \tag{24}$$

where $T_{\text{w}}$ = 0.167 s is the RoCoF window, which is moved in steps of 0.01 s. Each simulation is initialized until it reaches steady state, and samples whose pre-outage frequency lies outside 59.966 to 60.034 Hz are discarded, so that the labels are measured from a settled operating point. The units that remain at the contingency bus are excluded from $\mathcal{G}_m$ because their rotor speeds are measured at the disturbed bus itself, so that the labels describe the response of the generators that are connected to the disturbance through the network.

Because both labels are taken over the individual generators, they retain the effects of the heterogeneous governor dynamics and of the spatially nonuniform responses that the aggregated SFR model in Section II.B neglects.

The inputs of the NN-FP are the setpoints sent to the EMT model, whereas the pre-outage operating point is established by the EMT model itself. Because the DC power flow uses a fixed loss allocation, the setpoints and the actual losses are not balanced exactly, and the mismatch is shared by the governors of all units through their droop characteristics, in proportions that depend on the unit capacities, droop settings and governor limits. The realized outputs therefore differ slightly from the setpoints, and the pre-outage frequency settles at a value that differs slightly from $f_0$. Since the labels are measured from the realized response, this relation between the scheduled and the realized operating point is part of the mapping that the NN-FP learns.

*3) Structure and training of the NN-FP*: A fully connected NN is adopted as the frequency predictor, since it can learn the nonlinear relationship between the operating condition and the frequency response metrics without imposing an aggregated analytical response model. As shown in Fig. 2, the NN-FP takes $x$ as the input and returns $\text{RoCoF}^{\text{w}}$ and $\Delta f^{\text{w}}$ as the outputs. Each hidden layer applies an affine transformation followed by an elementwise rectified linear unit (ReLU) activation, and the output layer is linear. The forward mapping is given in (25) to (28).

$$z_1 = xW_1 + b_1 \tag{25}$$

$$z_m = a_{m-1}W_m + b_m, 2 \le m \le N_L \tag{26}$$

$$a_m = max(z_m, 0)\,,\ 1 \le m \le N_L \tag{27}$$

$$r_h = a_{N_L}W_{N_L+1} + b_{N_L+1} \tag{28}$$

where $W_m$ and $b_m$ denote the weight matrix and the bias vector of layer m, respectively; $N_L$ is the number of hidden layers; $a_m$ represents the activation output vector of layer m; $z_m$ is the pre-activation vector of layer m; $x$ denotes the input vector; and $r_h = [\text{RoCoF}^{\text{NN}}, \Delta f^{\text{NN}}]$ is the output vector.

The retained samples are divided into training, validation and test sets. The input and output scaling is fitted only to the training set. The NN-FP is trained on the training samples by minimizing the mean squared error (MSE) loss defined in (29).

$$\mathcal{L} = \frac{1}{n}\sum_{k=1}^{n} \left\| r_{h,k} - r_{true,k} \right\|_2^2 \tag{29}$$

where n is the number of training samples, $r_{\text{true,k}}$ is the label vector of sample k, and ‖·‖2 denotes the Euclidean norm.

The validation set is used to monitor the training, and the retained weights are specified for each test system in Section IV. The test samples, which are used neither for training nor for model selection, quantify the prediction errors and provide the empirical margins introduced in Section III.B.

### B. Online Stage: NN-FCOPF Formulation

*1) Exact reformulation of the NN-FP*: In the online stage, the trained NN-FP is embedded in the T-OPF as a set of constraints on the decision variables $P_i$, while the loads of the dispatch interval enter the input vector as fixed parameters. The affine mappings (25), (26) and (28) are linear in the decision variables and can be included directly. The ReLU activation (27) is nonlinear, but it can be represented exactly by the big-M method [22] with one binary variable per hidden neuron, as shown in (30) to (33) for neuron *j* of layer *m*.

$$a_{m,j} \le z_{m,j} - h_{m,j}^{l}\left(1 - B_{m,j}\right) \tag{30}$$

$$a_{m,j} \ge z_{m,j} \tag{31}$$

$$a_{m,j} \le h_{m,j}^{u} B_{m,j} \tag{32}$$

$$a_{m,j} \ge 0 \tag{33}$$

where $h_{m,j}^{u}$ and $h_{m,j}^{l}$ are the upper and lower bounds of the pre-activation $z_{m,j}$ over all admissible inputs, and $B_{m,j}$ is the corresponding binary variable, for every neuron *j* of layer *m*.

The reformulation is exact, so the outputs $\mathrm{RoCoF}^{\mathrm{NN}}$ and $\Delta f^{\mathrm{NN}}$ inside the optimization coincide with the predictions of the trained network at the same operating point.

*2) Additional constraints*: As illustrated in the online part of Fig. 2, additional constraints are imposed on the outputs and the inputs of the NN-FP. The predicted outputs are bounded by the frequency limits:

$$\mathrm{RoCoF}^{\mathrm{NN}} \le R_{\mathrm{lmt}} - \delta_R \tag{34}$$

$$\Delta f^{\mathrm{NN}} \le \Delta f_{\mathrm{lmt}} - \delta_f \tag{35}$$

where $R_{\mathrm{lmt}}$ and $\Delta f_{\mathrm{lmt}}$ are the RoCoF and FD limits, the RoCoF is expressed as a magnitude, and $\delta_R$ and $\delta_f$ are safety margins.

The optimizer tends to select operating points at which the NN-FP underpredicts the frequency excursion, because such points appear less costly under the embedded constraints. The limits in (34) and (35) are therefore tightened by the largest underprediction of the NN-FP on the test set $\mathcal{T}$, which is used neither for training nor for model selection:

$$\delta_R = \max\left\{0, \max_{n\in\mathcal{T}}\left(RoCoF_n - RoCoF_n^{NN}\right)\right\} \tag{36}$$

$$\delta_f = \max\left\{0, \max_{n\in\mathcal{T}}\left(\Delta f_n - \Delta f_n^{NN}\right)\right\} \tag{37}$$

where $\mathrm{RoCoF}_n$ and $\Delta f_n$ are the EMT labels of test sample *n*, and $\mathrm{RoCoF}_n^{\mathrm{NN}}$ and $\Delta f_n^{\mathrm{NN}}$ are the corresponding predictions. The outer maximum keeps the margins nonnegative.

The margins are empirical; they reflect the prediction error of the NN-FP on unseen operating conditions and are verified against EMT simulations in Section IV. A formal guarantee on the prediction error would require verification of the trained network [23], which is beyond the scope of this paper.

In addition, the generation components of the input vector are restricted to the range covered by the training data:

$$x^{G,min} \le x^{G} \le x^{G,max} \tag{38}$$

where $x^G$ collects the generation components of $x$, and $x^{G,min}$ and $x^{G,max}$ are the elementwise minimum and maximum of these components over the training samples. This constraint keeps each generation input within the range of values seen in training.

The NN-FCOPF minimizes (1) subject to (2) to (5), (25), (26), (28), (30) to (35) and (38), with the margins defined in (36) and (37); it is a mixed-integer quadratic program, which is solved for each dispatch interval. The objectives and constraints of the three models are summarized in Table I.

TABLE I
OBJECTIVES AND CONSTRAINTS OF T-OPF, L-FCOPF AND NN-FCOPF

| Model | Objective | Shared constraints | Unique constraints |
|---|---|---|---|
| T-OPF | | | None |
| L-FCOPF | (1) | (2) to (5) | (9), (22) |
| NN-FCOPF | | | (25), (26), (28), (30) to (35), (38) |

## IV. CASE STUDIES

This section verifies the effectiveness of the proposed NN-FCOPF through comparisons with the two benchmark models, T-OPF and L-FCOPF, on the IEEE 9-bus and 39-bus test systems.

### A. IEEE 9-Bus System

The topology of the modified IEEE 9-bus system is shown in Fig. 3, and the system parameters are given in [24]. The total generator outputs at buses 1, 2 and 3 in the base case are 72 MW, 163 MW and 85 MW, respectively. The generator at each bus is split into several identical units, namely two at bus 1, four at bus 2 and three at bus 3. The units at the same bus share the same rated capacity, governor and cost function and are dispatched with equal outputs, so that the output of each unit equals the bus total divided by the number of units. The base case loads at buses 5, 6 and 8 are 125 MW, 90 MW and 100 MW, respectively.

The parameters of the full-order SFR model in Fig. 1(a) are listed in Table II for each generator, where m = 1, 2, j = 1, 2, 3, 4 and p = 1, 2, 3 denote the unit indices at buses 1, 2 and 3. *H* is the inertia constant of each unit in seconds on its rating.

The T-OPF, L-FCOPF and NN-FCOPF are implemented for this system according to Table I. The three models share all settings except the frequency constraints. The following subsections describe the training and validation of the NN-FP, the implementation of the optimization models, and the EMT validation of the resulting dispatches.

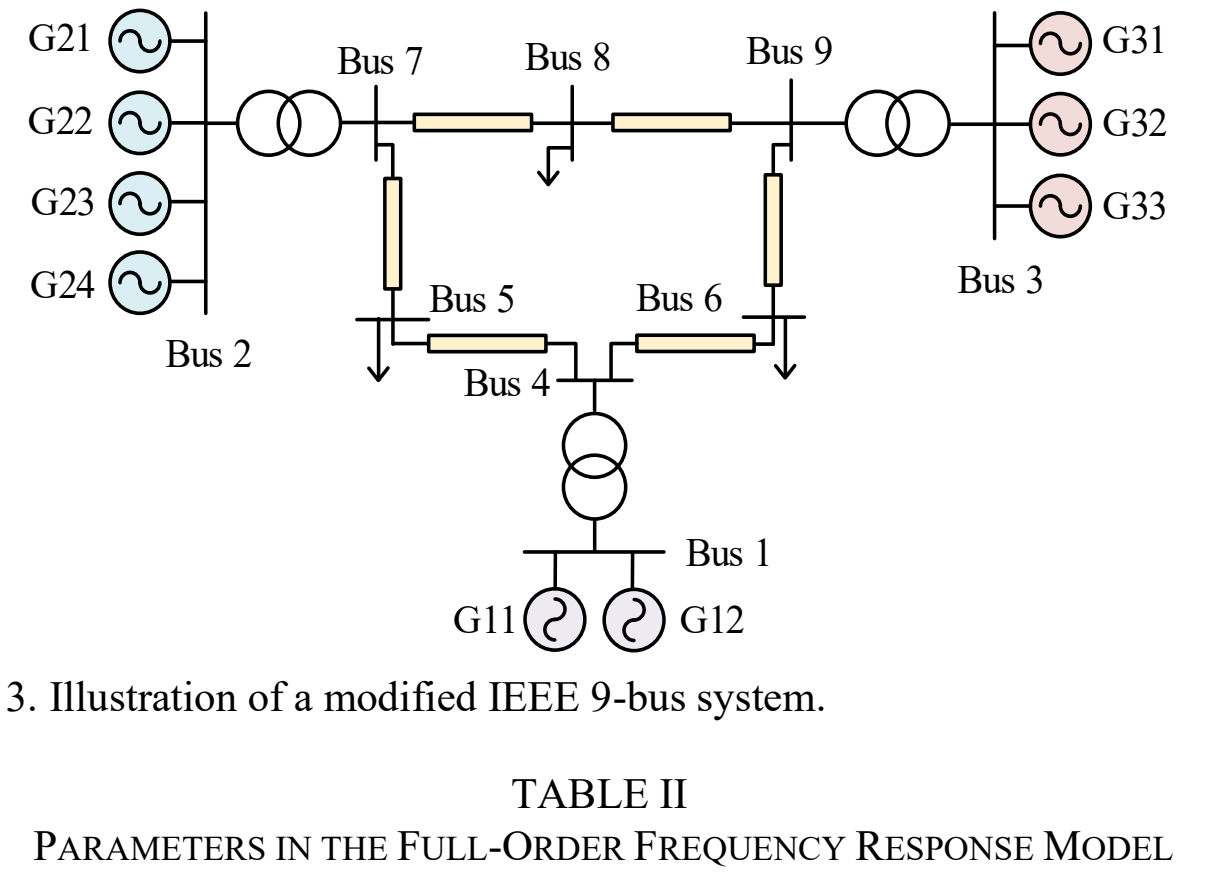

Fig. 3. Illustration of a modified IEEE 9-bus system.

TABLE II
PARAMETERS IN THE FULL-ORDER FREQUENCY RESPONSE MODEL

| Gen | $T_1$/s | $T_2$/s | $T_3$/s | $T_4$/s | $T_5$/s | $F$ | $K$ | $R$ | $H$ |
|---|---|---|---|---|---|---|---|---|---|
| G1*m* | 0.15 | 0.05 | 0.3 | 0.26 | 8 | 0.27 | 1 | 0.05 | 7 |
| G2*j* | 0.1 | 0 | 0.259 | 0.1 | 10 | 0.272 | 1 | 0.05 | 4 |
| G3*p* | 0.083 | 0 | 0.2 | 0.05 | 5 | 0.28 | 1 | 0.05 | 3.117 |

*1) Training of NN-FP*: The accuracy of the NN-FP directly determines the quality of the NN-FCOPF dispatch, so a reliable training dataset is essential. To this end, an EMT model of the modified IEEE 9-bus system is established in PSCAD/EMTDC with the WSCC Type G governor [25] and the IEEE AC5C exciter [26]. Although all generators use the same governor structure, which corresponds to a system dominated by reheat steam turbines, their parameters, such as the time constants in Table II, differ among the buses. This diversity provides a realistic representation of the different response characteristics of the individual units during frequency disturbances. In the EMT model, the dispatch setpoint of each unit enters the reference input of its governor.

The outage of unit G11, one of the two units at bus 1, is considered as the contingency. The contingency is selected by rating and inertia. The units at bus 1 have the largest rated capacity and inertia constant in the system, so the outage of one of them removes the largest share of kinetic energy among the single unit outages, independently of the dispatch. The operating points are generated by the two routes of Section III.A, namely T-OPF solutions with randomly varied bus loads, whose total ranges from 72% to 135% of the base case load, and Latin hypercube samples. Each operating point is simulated under this outage, the samples that pass the pre-outage frequency screening are retained, and the retained samples are divided into training, validation and test sets. The labels are the worst RoCoF and the worst FD over the rotor speed frequencies of the units at buses 2 and 3, which form the monitored set $\mathcal{G}_m$, and each sample pairs these two labels with the corresponding generator outputs and loads. The same EMT model also serves as the reference for validating the dispatches of the three optimization models.

*2) Validation of NN-FP*: The NN-FP for the IEEE 9-bus system has 6 inputs, namely the total generation at the three generator buses and the three loads, one hidden layer with 32 ReLU neurons, and two outputs. It is trained with the Adam optimizer at a learning rate of 0.005, with a batch size of 32 for 200 epochs. Fig. 4 shows the training and validation loss curves and compares the predictions with the EMT labels on the unseen test samples. Table III summarizes the mean absolute error (MAE), the root mean squared error (RMSE), the coefficient of determination and the empirical safety margins on the test set. The MAE is 0.0013 Hz/s for RoCoF and 0.0034 Hz for FD, and the coefficients of determination exceed 0.999 for both outputs. The largest underpredictions on the test set, which are used as the margins $\delta_R$ and $\delta_f$ in (34) and (35), are 0.0050 Hz/s and 0.0132 Hz, respectively.

These results show that the NN-FP predicts both indicators accurately on unseen operating conditions and is therefore suitable for embedding into the dispatch model.

*3) Development of NN-FCOPF*:

The trained NN-FP is reformulated according to Section III.B and embedded into the NN-FCOPF. The proposed model and the two benchmark models are implemented in Python 3.9 and solved with Gurobi. All models share the same parameters. In L-FCOPF and NN-FCOPF, the RoCoF limit is 0.5 Hz/s [17]. The FD limit is 0.5 Hz in this study. The RoCoF limit applies to the RoCoF magnitude in (34). All problems are solved on a computer with an Intel Core i7-12700 CPU at 2.1 GHz and 32 GB of RAM. The training range constraint (38) is not active in any of the 24 hours, i.e., every generation input stays within its training range.

*4) Validation of NN-FCOPF*: The dispatches of the three models are validated by EMT simulation over a full daily cycle. The 24-h load profile derived from the demand data of the Electric Reliability Council of Texas (ERCOT), taken from [27], is applied to the three models. The original profile is shown in Fig. 5.

The contingency is the outage of unit G11 at $t = 0$ s. The three models are solved for each hour, and the resulting generator setpoints are applied as references in the PSCAD/EMTDC model, where the outage is simulated and the worst RoCoF and FD are recorded. Fig. 6 compares the FD and the RoCoF obtained from the EMT simulations for the three dispatches over the 24 hours, and Fig. 7 shows the prediction errors of the L-FCOPF and NN-FCOPF with respect to the EMT results. The T-OPF is excluded from Fig. 7 because it does not predict the frequency response.

As shown in Fig. 6, the NN-FCOPF satisfies both limits in all 24 hours, with a largest FD of 0.496 Hz and a largest RoCoF of 0.420 Hz/s. In contrast, the T-OPF violates the FD limit in 11 hours, with a largest FD of 0.802 Hz, and the RoCoF limit in 3 hours. The L-FCOPF performs only marginally better than T-OPF.

TABLE III
TEST ACCURACY AND SAFETY MARGINS OF THE 9-BUS NN-FP

| Output | MAE | RMSE | $R^2$ | $\delta$ |
|---|---|---|---|---|
| RoCoF (Hz/s) | 0.0013 | 0.0018 | 0.9998 | 0.0050 |
| $\Delta f$ (Hz) | 0.0034 | 0.0047 | 0.9996 | 0.0132 |

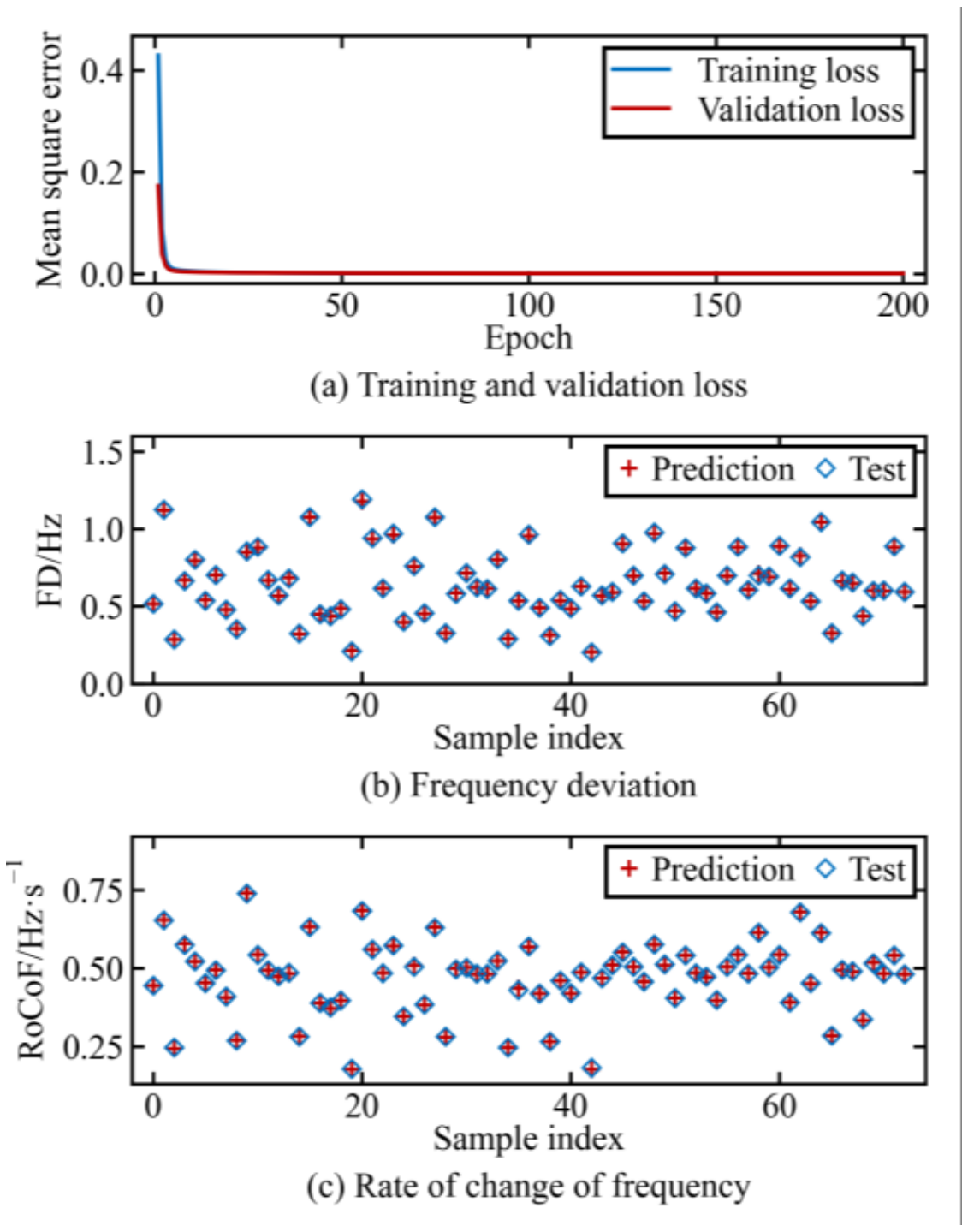

Fig. 4. Validation of the NN-FP for the IEEE 9-bus system: (a) training and validation loss curves, (b) frequency deviation predictions and EMT labels, and (c) RoCoF predictions and EMT labels.

In this system, the FD is the binding frequency requirement as the FD constraint of the NN-FCOPF is active in 13 hours, whereas its RoCoF constraint is never active. However, in the L-FCOPF the frequency constraints modify the dispatch only in Hours 7 to 10, where the SFR model predicts an FD of 0.50 Hz while the EMT simulation yields 0.67 to 0.79 Hz. In the remaining hours, the SFR model regards the T-OPF dispatch as secure although the FD exceeds 0.5 Hz in seven of them. Consequently, the L-FCOPF also violates the FD limit in 11 hours.

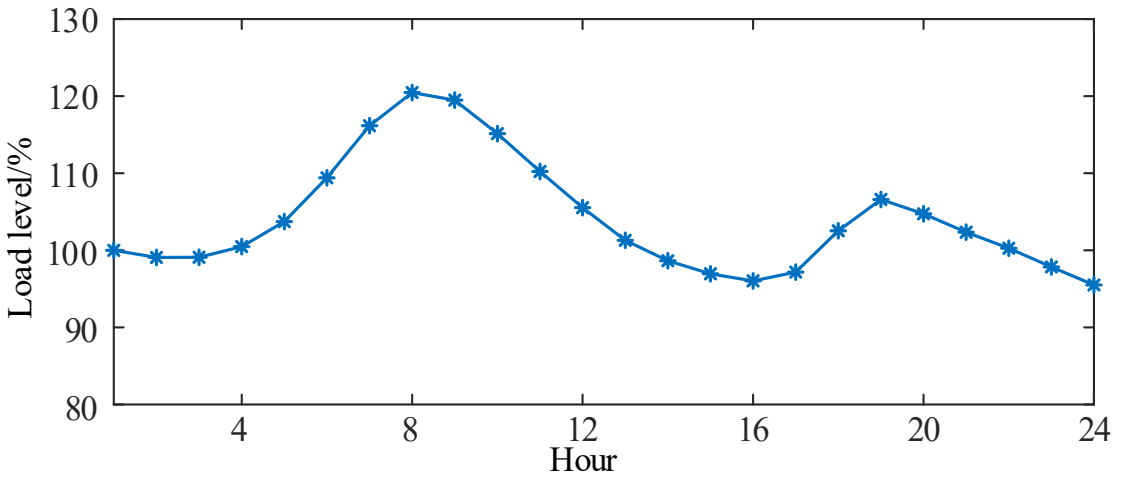

Fig. 5. 24-hour load profile from ERCOT.

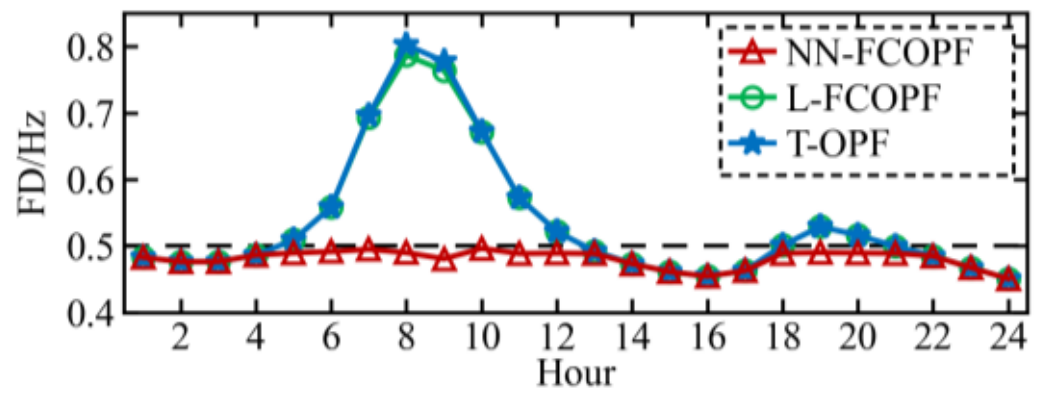

(a) Frequency deviation

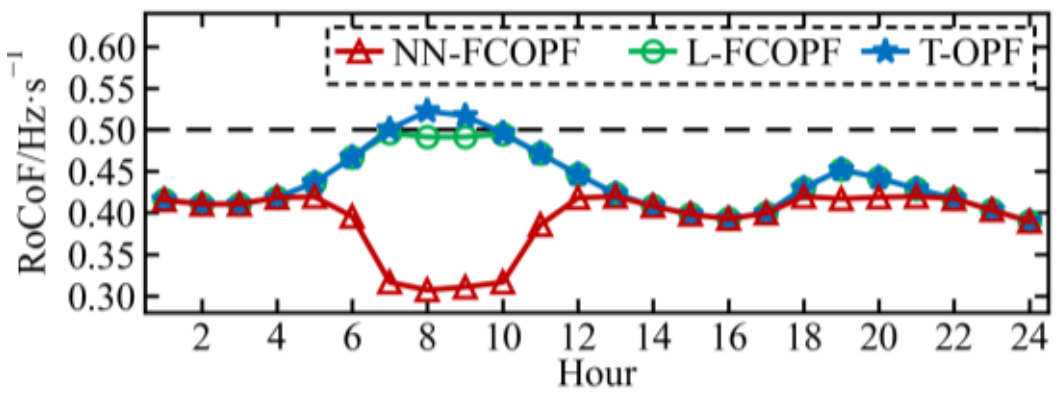

(b) Rate of change of frequency

Fig. 6. Comparison of (a) FD and (b) RoCoF over 24 hours obtained from EMT simulations for the IEEE 9-bus system under T-OPF, L-FCOPF, and NN-FCOPF.

The NN-FCOPF increases the 24-h operating cost from \$133,764.03 under the T-OPF to \$133,781.20, i.e., by only 0.013%; the 24-h cost of the L-FCOPF is \$133,764.08, so the three models differ in cost by less than 0.013%.

Over the whole day, the NN-FCOPF is therefore the only one of the three models that respects both frequency limits, and it does so at essentially the same cost.

The prediction errors in Fig. 7 are defined in (39), with the EMT results as the reference.

$$e = y_d - y_s \tag{39}$$

where $y_d$ is the value predicted by the optimization model and $y_s$ is the value measured in the EMT simulation. A negative error means that the model underestimates the frequency excursion and therefore overestimates the frequency security.

As shown in Fig. 7, the NN-FCOPF predicts both indicators accurately, with FD errors between −0.009 and +0.006 Hz and RoCoF errors between −0.002 and +0.008 Hz/s. The SFR model of the L-FCOPF underestimates the FD by 0.08 to 0.29 Hz and the RoCoF by 0.063 to 0.075 Hz/s in every hour, i.e., its errors are always on the unsafe side. Since the EMT RoCoF of the L-FCOPF stays below the limit in this system, the FD error of the SFR model is responsible for the violations of the L-FCOPF in Fig. 6. This error stems from the simplified governor representation: as shown below for Hour 8, the frequencies at the three generator buses differ by less than 0.002 Hz at the nadir, so the locational effect on the FD is negligible in this system.

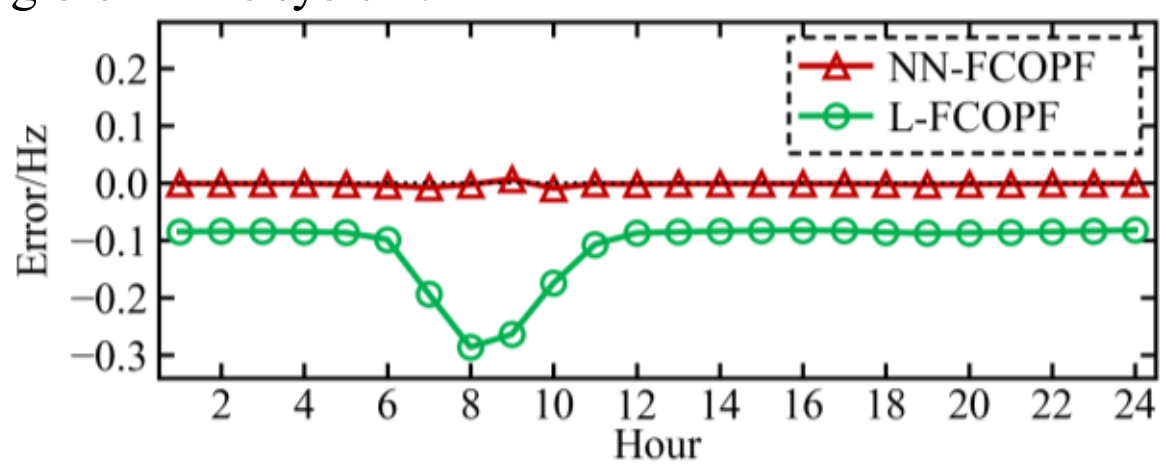

(a) FD prediction error

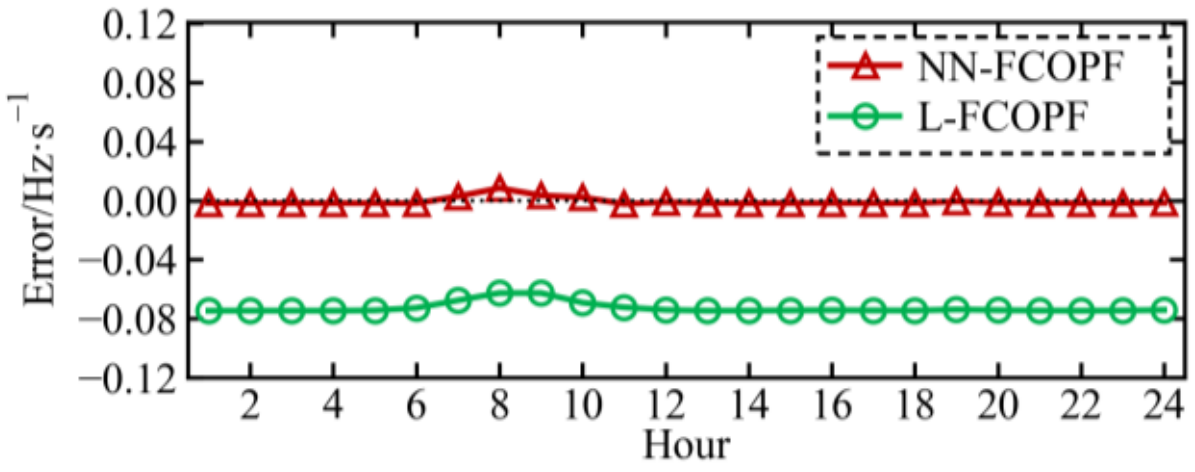

(b) RoCoF prediction error

Fig. 7. Prediction errors of (a) FD and (b) RoCoF of L-FCOPF and NN-FCOPF with respect to the EMT simulation results for the IEEE 9-bus system.

Hour 8, the daily peak, is examined in detail. Table IV summarizes the optimization and EMT validation results, and Fig. 8 compares the corresponding EMT frequency responses.

In this hour, the T-OPF violates both limits, with an FD of 0.80 Hz and a RoCoF of 0.522 Hz/s. The FD constraint of the L-FCOPF is active and reduces the output of the units at bus 1, but the SFR model underestimates the FD by 36.43%, so the EMT simulation still yields an FD of 0.79 Hz. The NN-FCOPF reduces the output of each unit at bus 1 to 29.9 MW, compared with 49.6 MW in the T-OPF and 46.6 MW in the L-FCOPF, and both limits are satisfied, with an FD of 0.49 Hz and a RoCoF of 0.307 Hz/s. The absolute relative prediction errors of the NN-FCOPF are 0.63% for the FD and 2.72% for the RoCoF, compared with 36.43% and 12.76% for the L-FCOPF.

The traces in Fig. 8 show the subsequent recovery and oscillation without changing the reported extrema.

TABLE IV
RESULTS OF T-OPF, L-FCOPF AND NN-FCOPF AT HOUR 8

| Model | | T-OPF | L-FCOPF | NN-FCOPF |
|---|---|---|---|---|
| OPF | Hourly Cost (\$) | 6036.5 | 6036.5 | 6043.8 |
| | Solve Time (s) | 0.001 | 0.001 | 0.089 |
| | RoCoF (Hz/s) (error, %) | N/A | 0.429 (12.76) | 0.315 (2.72) |
| | $\Delta f$ (Hz) (error, %) | N/A | 0.500 (36.43) | 0.487 (0.63) |
| Simulation | RoCoF (Hz/s) | 0.522 | 0.491 | 0.307 |
| | $\Delta f$ (Hz) | 0.802 | 0.786 | 0.490 |

Fig. 9 compares the frequencies recorded at generator buses 1 to 3 for the NN-FCOPF dispatch at Hour 8. The trajectories differ among the locations only during the initial oscillations: the nadirs at the three buses lie within 0.0015 Hz of each other (59.5101 to 59.5116 Hz), and the RoCoF extrema range from 0.271 to 0.307 Hz/s. The training labels take the extrema over the generators at buses 2 and 3, the buses other than the contingency bus, and bus 2 determines both extrema in this case, with an FD of 0.4899 Hz and a RoCoF magnitude of 0.3071 Hz/s. The trace at bus 1 is shown for reference only, since bus 1 is the contingency bus and is not included in $\mathcal{G}_m$. In this system, the spatial variation is small, and the advantage of the NN-FP over the SFR model comes from its representation of the governor dynamics.

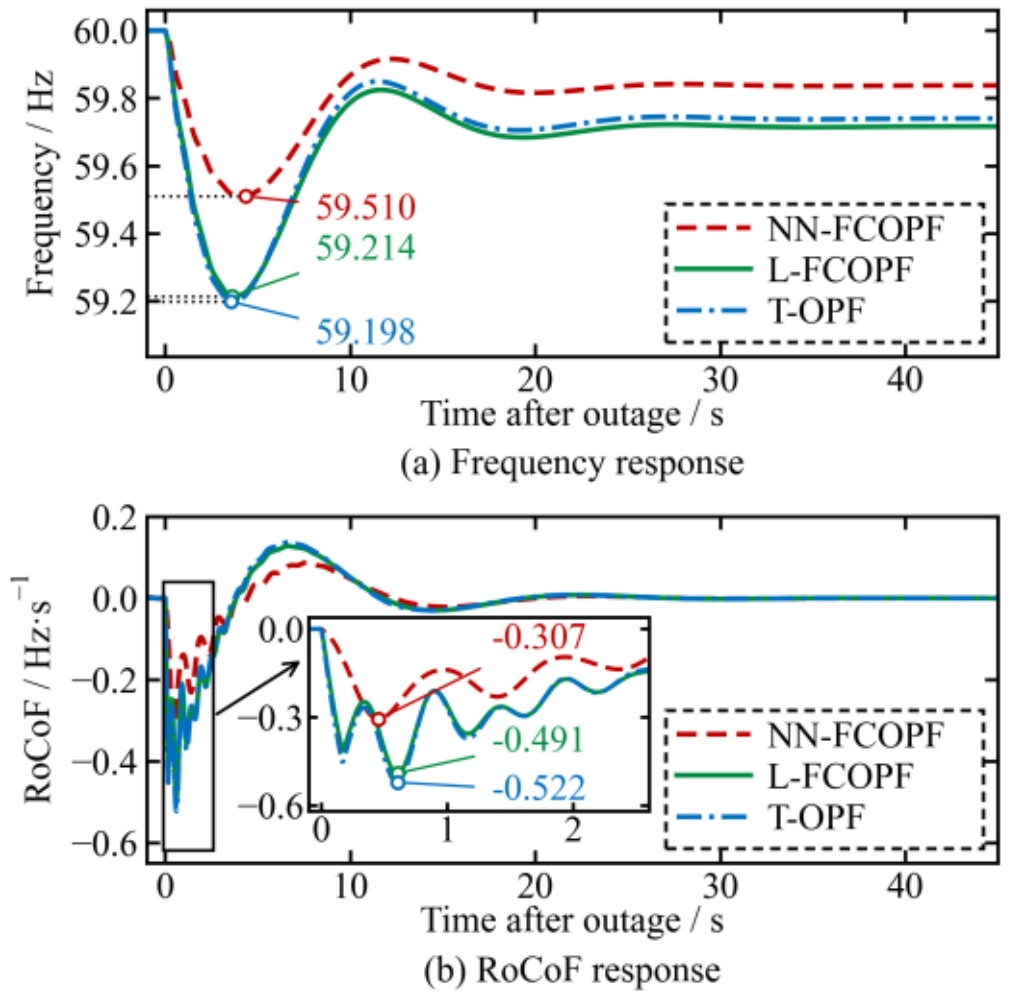

Fig. 8. EMT responses at Hour 8 under T-OPF, L-FCOPF, and NN-FCOPF: (a) frequency at the generator bus with the largest FD and (b) RoCoF at the generator bus with the largest RoCoF magnitude.

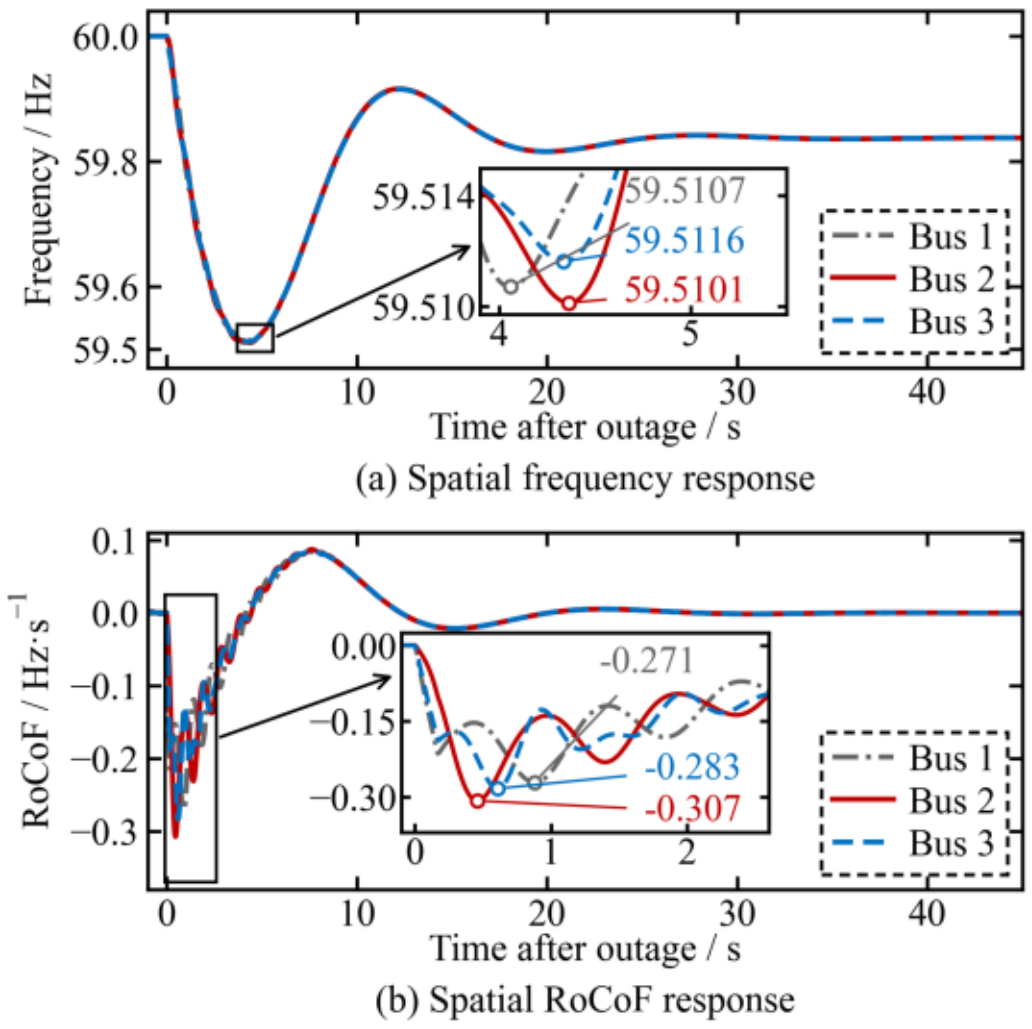

Fig. 9. Spatial variation of the frequency at the generator buses at Hour 8 under NN-FCOPF: (a) spatial frequency response and (b) RoCoF.

To evaluate the safety margins in (34) and (35), the NN-FCOPF is also solved with $\delta_R = \delta_f = 0$, i.e., with the nominal limits, for the same 24-h load profile. As listed in Table V, the dispatch without margins satisfies the embedded constraints, and the predicted FD equals the limit of 0.5 Hz in 10 hours. The EMT simulation, however, shows FD violations in 9 hours with a largest FD of 0.509 Hz, because the NN-FP slightly underpredicts the FD at these operating points. The violations are of the same order as the prediction error of the NN-FP, which is exactly what the margins are designed to cover. With the test set margins, both limits are satisfied in all 24 hours, and the 24-h cost increases from \$133,779.37 to \$133,781.20, i.e., by only 0.0014% relative to the dispatch without margins.

In summary, the case studies on the IEEE 9-bus system show that the NN-FCOPF keeps both frequency indicators within their limits in every hour, whereas the T-OPF and the L-FCOPF violate the FD limit in 11 hours each. The improvement is obtained at a negligible increase in operating cost. In this system the improvement comes from the representation of the governor dynamics, since the spatial variation of the frequency is small.

TABLE V
Effect of the Safety Margins on the IEEE 9-Bus System

| Setting | Without margins | With margins |
|---|---|---|
| RoCoF bound in OPF (Hz/s) | 0.500 | 0.495 |
| $\Delta f$ bound in OPF (Hz) | 0.500 | 0.487 |
| Largest EMT RoCoF (Hz/s) | 0.430 | 0.420 |
| Largest EMT $\Delta f$ (Hz) | 0.509 | 0.496 |
| Hours violating FD limit | 9 | 0 |
| Hours violating RoCoF limit | 0 | 0 |

## B. IEEE 39-Bus System

To examine the scalability of the proposed framework, case studies are conducted on the IEEE 39-bus system shown in Fig. 10, which has 10 synchronous generators and whose parameters are given in [28]. The governor and turbine parameters of Table II are also used in this system: the generators at buses 30 to 32, 33 to 37 and 38 to 39 use the parameters of G2, G3 and G1 in the modified IEEE 9-bus system, with inertia constants of 4 s, 3.117 s and 7 s, respectively. The generator at bus 38 has the largest rated capacity in the system, 2000 MVA, with an inertia constant of 7 s, and its outage is the specified contingency, chosen for the same reason as in the 9-bus system.

The NN-FP has 29 inputs, namely 10 generator outputs and 19 loads. The operating points are generated by T-OPF solutions and Latin hypercube sampling as in Section III.A, simulated under the outage of the generator at bus 38, screened and divided into training, validation and test sets in the same way as for the 9-bus system. The labels are taken over the frequencies at all generator buses other than bus 38, the bus where the contingency occurs, so that $\mathcal{G}_m$ contains all remaining generators. The NN-FP again has one hidden layer with 32 ReLU neurons and is trained with the Adam optimizer at a learning rate of 0.005, with a batch size of 32 for 1000 epochs; the weights of the epoch with the lowest validation loss are retained. On the test set, the MAE is 0.0185 Hz/s for RoCoF and 0.0133 Hz for FD, and the resulting margins are 0.0280 Hz/s and 0.0388 Hz. The training range constraint (38) is not active in any hour.

Fig. 11 compares the FD and the RoCoF obtained from the EMT simulations for the three models, Fig. 12 shows the prediction errors of the L-FCOPF and NN-FCOPF, and Table VI summarizes the 24-h results.

Under the T-OPF dispatch, the outage leads to a RoCoF of 1.35 to 1.57 Hz/s and an FD of 0.82 to 1.28 Hz, which violate both limits in all hours. The RoCoF constraint of the L-FCOPF is active in all hours, so the SFR model predicts a RoCoF of exactly 0.5 Hz/s. The EMT simulation, however, yields a RoCoF of 0.92 to 1.04 Hz/s and an FD of 0.64 to 0.79 Hz, so the L-FCOPF also violates both limits in all hours. The RoCoF constraint of the NN-FCOPF is active in all hours and its FD constraint in Hour 8, and the EMT simulation confirms that both limits are satisfied in all hours, with a largest RoCoF of 0.493 Hz/s and a largest FD of 0.453 Hz.

As shown in Fig. 12, the NN-FP predicts the RoCoF with errors between −0.021 and −0.002 Hz/s and the FD with errors between −0.001 and +0.008 Hz. Since the magnitude of every RoCoF error is smaller than the margin $\delta_R = 0.028$ Hz/s, the measured RoCoF stays below 0.5 Hz/s although the RoCoF constraint is active in every hour. The SFR model, in contrast, underestimates the RoCoF by 0.42 to 0.54 Hz/s and the FD by 0.17 to 0.32 Hz.

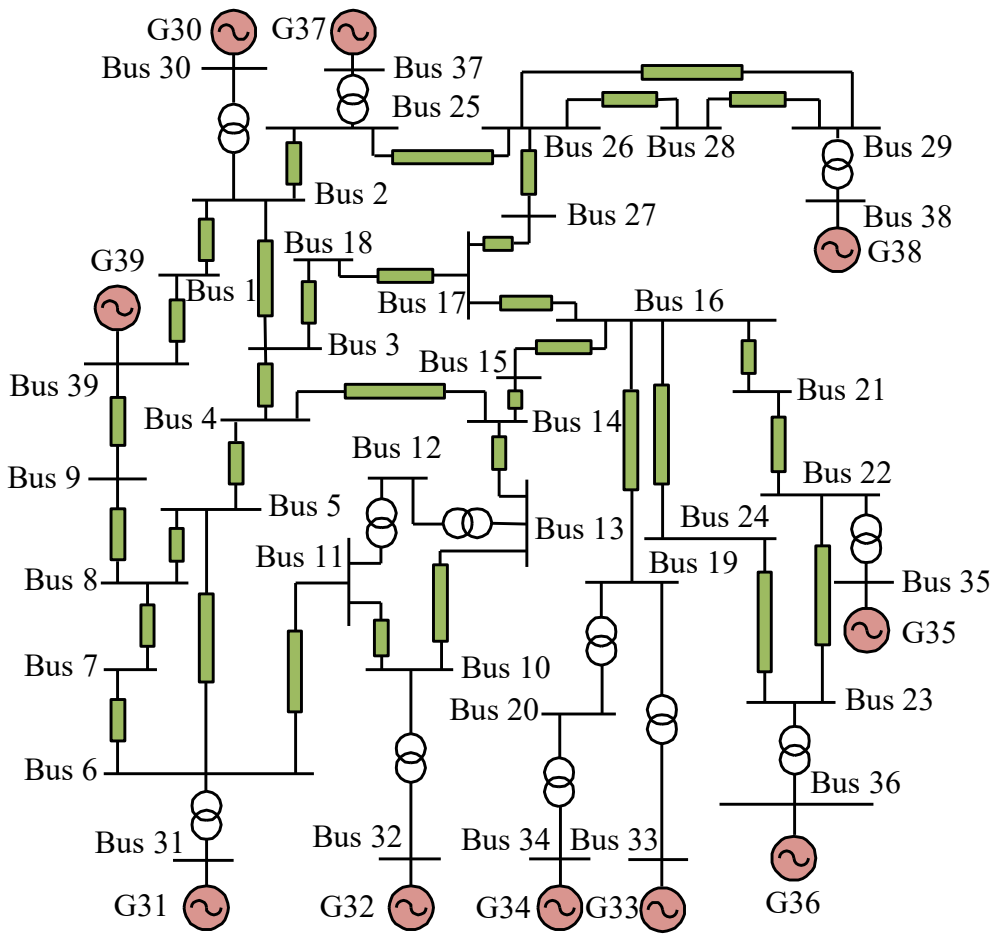

Fig. 10. Topology of the IEEE 39-bus system.

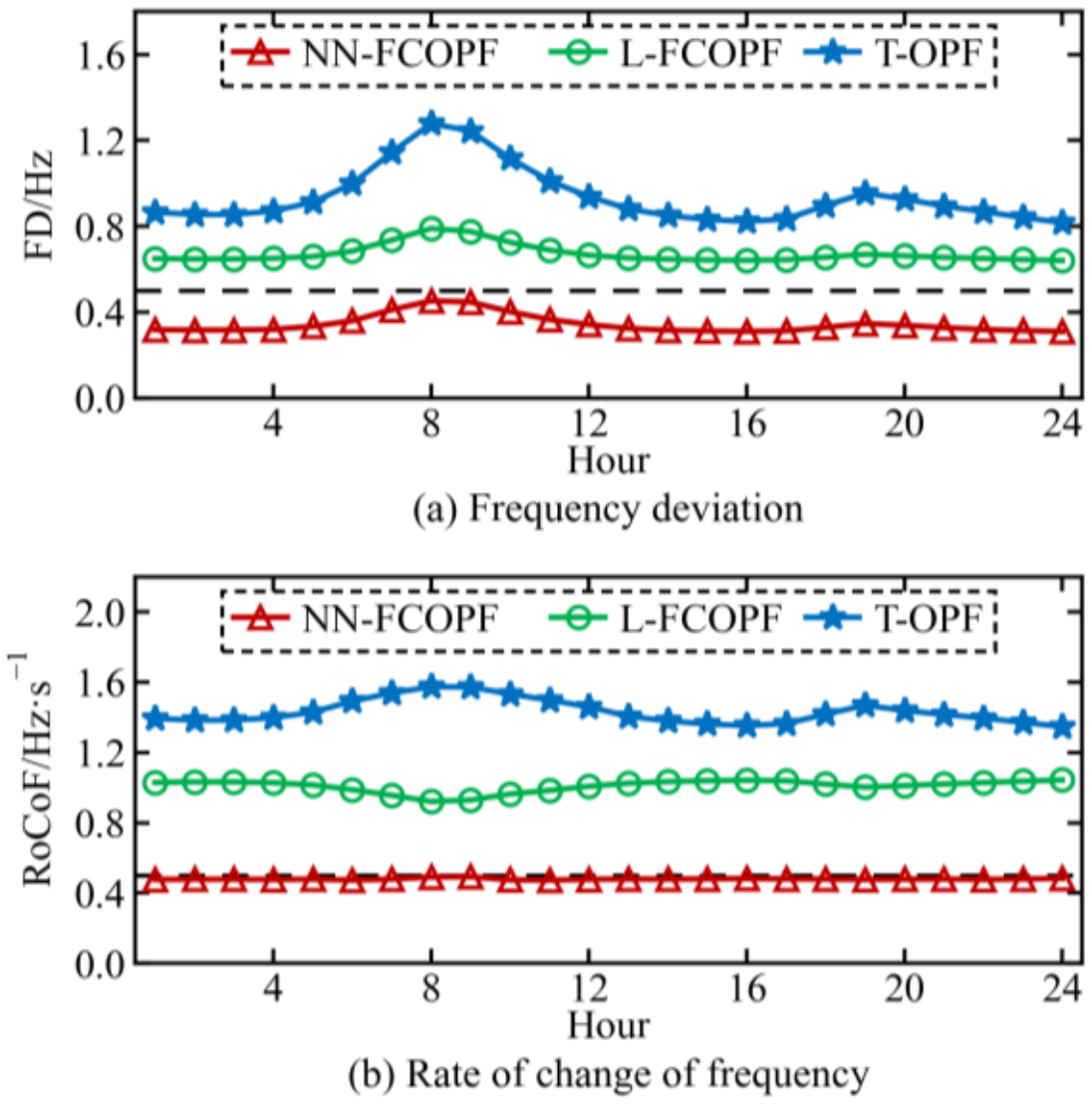

Fig. 11. Comparison of (a) FD and (b) RoCoF over 24 hours obtained from EMT simulations for the IEEE 39-bus system under T-OPF, L-FCOPF, and NN-FCOPF.

The reason for the large RoCoF error of the L-FCOPF is examined in Fig. 13, which shows the rotor speed frequencies and the RoCoF of all remaining generators and of the COI, the latter being the inertia weighted average of the rotor speeds according to (6), after the outage at Hour 8 of the L-FCOPF dispatch. In Fig. 13(a), the generator frequencies spread around the COI frequency during the first second and then decline together, so that the lowest generator frequency, 59.2103 Hz at bus 34, differs from the COI nadir of 59.2172 Hz by only 0.007 Hz. The corresponding FD values, 0.79 Hz for the worst generator and 0.78 Hz for the COI, therefore almost coincide, and the FD error of the SFR model in this hour stems from its simplified governor representation, since the locational effect on the FD is small.

In Fig. 13(b), by contrast, the generator at bus 37 swings against the rest of the system immediately after the outage and reaches a RoCoF of 0.92 Hz/s, whereas the RoCoF of the COI frequency is 0.54 Hz/s. Over the 24 hours, the COI RoCoF measured in the EMT simulation, 0.52 to 0.54 Hz/s, is close to the 0.5 Hz/s predicted by the SFR model, whereas the worst generator RoCoF, 0.92 to 1.04 Hz/s, is about twice as large. The SFR model therefore describes the COI RoCoF reasonably well, but the frequency security of the individual generators requires the worst generator metrics that the NN-FP is trained to predict.

TABLE VI
24-HOUR RESULTS OF THREE FRAMEWORKS FOR THE IEEE 39-BUS SYSTEM

| Model | T-OPF | L-FCOPF | NN-FCOPF |
|---|---|---|---|
| 24-h cost ($) | 11,541,905 | 11,672,247 | 12,191,444 |
| Cost increase vs. T-OPF | - | 1.13% | 5.63% |
| Max. solve time per hour (s) | 0.003 | 0.004 | 0.234 |
| EMT RoCoF (Hz/s) | 1.346–1.574 | 0.921–1.044 | 0.474–0.493 |
| EMT $\Delta f$ (Hz) | 0.82–1.28 | 0.64–0.79 | 0.31–0.45 |
| Hours violating RoCoF / FD limit | 24 / 24 | 24 / 24 | 0 / 0 |

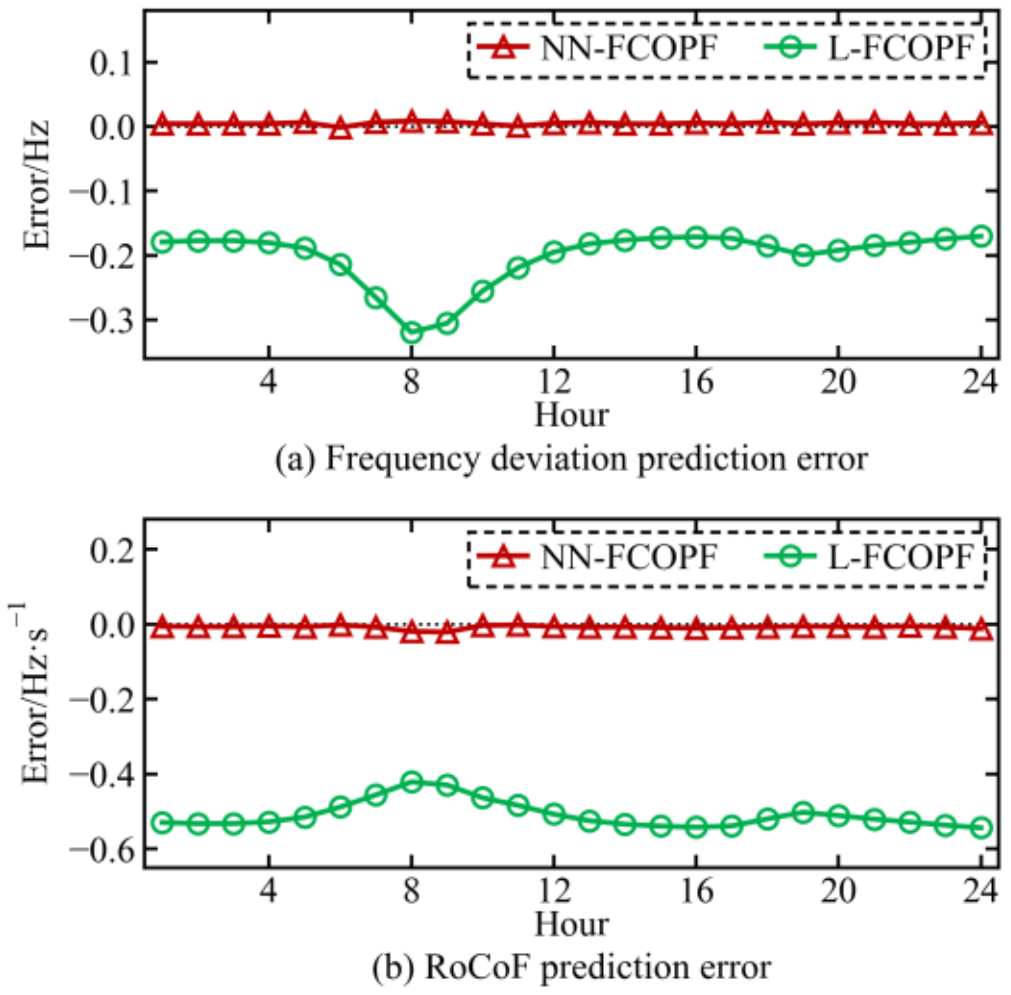

Fig. 12. Prediction errors of (a) FD and (b) RoCoF of L-FCOPF and NN-FCOPF with respect to the EMT simulations for the IEEE 39-bus system.

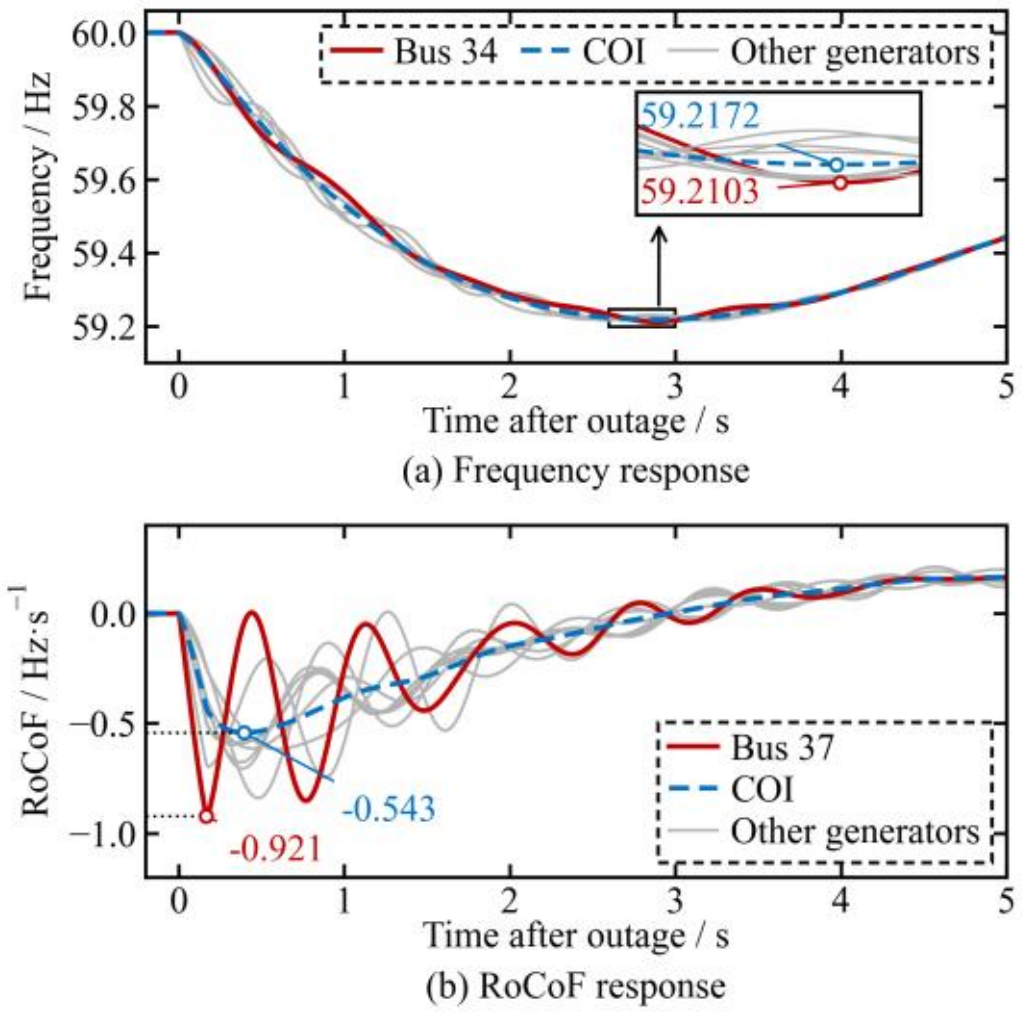

Fig. 13. Comparison of the COI and individual generator responses for the IEEE 39-bus system under the L-FCOPF dispatch at Hour 8: (a) frequency and (b) RoCoF after the outage of the generator at bus 38.

In terms of cost, the NN-FCOPF increases the 24-h cost by 5.63% relative to the T-OPF, compared with 1.13% for the L-FCOPF, which however does not achieve frequency security. The increase is larger than in the 9-bus system because the RoCoF constraint is active in every hour, so the output of the generator at bus 38 is reduced and replaced by more expensive generation throughout the day, whereas the frequency constraints modify the 9-bus dispatch in only part of the day.

The case studies on the two systems demonstrate that the NN-FP captures the governor dynamics, which dominate the FD error of the linearized model in the 9-bus system, and the locational RoCoF variation, which dominates its RoCoF error in the 39-bus system. Consequently, the NN-FCOPF satisfies both frequency limits in every hour of both systems, whereas the T-OPF and the L-FCOPF violate at least one limit in 11 hours or more. The results of the two systems also show that the binding frequency requirement depends on the system, namely the FD in the 9-bus system and the RoCoF in the 39-bus system, which confirms the need to predict both indicators.

## V. Conclusions

This paper proposes NN-FP, a neural network trained on EMT simulations to predict the worst RoCoF and FD among the monitored generators, which are the generators at all buses other than the contingency bus, and embeds it into real-time generation dispatch. The limits are tightened by the largest test set underprediction of the NN-FP. EMT validation of 24-hour dispatches on the modified IEEE 9-bus and 39-bus systems shows that the proposed NN-FCOPF satisfies both the RoCoF and FD limits in all hourly cases for the specified contingency. In contrast, the T-OPF and the L-FCOPF violate the limits in some hours. The results also show that the binding frequency requirement depends on the system, which confirms the need to predict both indicators.

## Acknowledgment

This research was supported in part by the National Science Foundation (NSF) under Grant 2337598, in part by the NSF AI Institute for Advances in Optimization (Award 2112533) and the Georgia Tech Strategic Energy Institute. Any opinions, findings, and conclusions or recommendations expressed in this paper are those of the authors and do not necessarily reflect the views of the NSF.